# Reasons behind the Water Crisis and its Potential Health Outcomes

Md. Galib Ishraq Emran, Rhidi Barma, Akram Hussain Khan, and Mrinmoy Roy

**ABSTRACT**

**Globally, the water crisis has become a significant problem that affects developing and industrialized nations. Water shortage can harm public health by increasing the chance of contracting water-borne diseases, dehydration, and malnutrition. This study aims to examine the causes of the water problem and its likely effects on human health. The study scrutinizes the reasons behind the water crisis, including population increase, climate change, and inefficient water management techniques. The results of a lack of water on human health, such as the spread of infectious diseases, a higher risk of starvation and dehydration, and psychological stress, are also concealed in the study. The research further suggests several ways to deal with the water situation and lessen its potential outcomes on human health. These remedies include enhanced sanitation and hygiene procedures, water management, and conservation techniques like rainwater gathering and wastewater recycling.**

**Keywords:** Climate change, Disease, Pollution, Sanitation, Water crisis.

**Md. Galib Ishraq Emran**
Department of Environmental Sciences, Jahangirnagar University, Savar, Dhaka, Bangladesh.
(e-mail: mdgalibishraqemran@gmail.com)

**Rhidi Barma**
Department of Environmental Sciences, Jahangirnagar University, Savar, Dhaka, Bangladesh.
(e-mail: rhidibarma003@gmail.com)

**Akram Hussain Khan**
Department of Statistics, Jahangirnagar University, Savar, Dhaka, Bangladesh.
(e-mail: akram45.ju@gmail.com)

**Mrinmoy Roy ***
Clinical Informatics Department, Ascension Illinois, Chicago, USA.
(e-mail: mrinmoy.cs10@gmail.com)

***Corresponding Author***

## I. INTRODUCTION

The ratio of land and water on the entire planet Earth is 1:2, and about 97.4% of the water belongs to the ocean and is too saline to drink or use without treatment (William & Henry J. Vaux, 2007). The majority of the remaining water is likewise unavailable because it is locked up in polar ice or glaciers. With this large portion of inaccessible water, humans and other terrestrial life must subsist on less than 1% (0.6%) of freshwater (William & Henry J. Vaux, 2007).

According to the United Nations, over 2 billion people lack access to clean drinking water, and over 4 billion lack access to properly run sanitary facilities (UN, 2023). The problem is particularly worse in rural parts of Africa and Asia, where, in many villages, the citizens have no access to sanitation or fresh water. Due to this condition, the population is forced to drink polluted water. They also use the polluted water regularly for things like taking a bath and cooking food. This lack of access to clean water and sanitation can lead to diseases and other health problems. How dangerous and emerging the water-based diseases are? How harmful are they to humans?

This unsanitary activity has a major negative influence on health. Waterborne diseases, including cholera, typhoid, and diarrhea, kill a lot of people annually, as children under the age of five experience 1.7 billion bouts of diarrhea, which results in 3 million fatalities. In addition, there are 11 million cases of typhoid fever and 117,000 fatalities as a result. About 88% of deaths related to diarrhea are brought on as a result of unsafe water, poor sanitation, and insufficient hygiene measures (CDC, 2023). In addition, exposure to contaminated water sources can result in chronic illnesses including fluorosis, arsenicosis, and other problems linked to water. These issues may have a long-term detrimental effect on their health and well-being (Robert & Jamie, 2008).

Using the limited freshwater resource, humankind has to do everything from drinking, irrigating, cooking, bathing, producing and manufacturing, bathing, and flushing toilets. As a result, since the 1950s, the worldwide water crisis has risen, yet freshwater supplies have been decreasing day by day (Debaere & Amanda, 2017). Half a billion people live in water-stressed or water-scarce countries, which is expected to increase to three billion by 2025 due to population growth (William & Henry J. Vaux, 2007). What will happen to all the people living in a water shortage for the rest of their lives? What effects does the water crisis have on human health? Due to lack of sufficient drinking water, people are suffering from cardiovascular diseases (Kakkat et al., 2023).

There is an urgent need to address the worldwide water problem and its effects on health, according to research from a number of sources. And what can be done to mitigate these problems?

According to studies, having access to clean water and sanitary facilities can lower the prevalence of waterborne illnesses and enhance general health outcome (Roy et al., 2023). For instance, research published in the Lancet Global Health revealed that better sanitation and water sources might reduce the global illness burden by up to 9.1% (Prüss-üstün & Carlos, 2007).

The study aimed to investigate the reasons behind the water crisis and its potential health outcomes. To do so, we comprehensively reviewed the literature from various sources. By researching this critical topic, we seek to increase awareness, support efforts to alleviate the water crisis, and enhance public health.

## II. Water Crisis

The other name of water is life. It is unthinkable to go through a day without water. Water is crucial for the survival of all organisms. Hence, water is called the blood of nature. A human being cannot survive without blood, and nature cannot sustain itself without water. Therefore, water is an essential element of the environment without which the environment will perish.

A water crisis is a situation in which the availability of potable, uncontaminated water in an area is less than the needs of that area. It is also known as a health crisis. Nearly one million people die each year from the diseases associated with water, sanitation, and hygiene (UN, 2023). It could be prevented if everyone had access to safe water and sanitation.

The water crisis is a global concern nowadays. It is created due to the unavailability of safe water according to demand. The water consumption rate varies from country to country. The per capita water use varies from country to country. From the AQUASTAT-FAO's information system on water and agriculture (2003), it is found that the per capita water in Mali is four cubic meters per year. In contrast, the per capita water use in the United States of America is 215 cubic meters per year. The per capita water use of the United States of America is 50 times higher than in Mali. And the per capita water use of other countries also varies from each other. For example, France consumes 106 cubic meters per year, Egypt consumes 77 cubic meters per year, India consumes 52 cubic meters per year, and China consumes 32 cubic meters per year.

According to the World Water Vision Report, there is a water crisis today. However, the issue is not about having too little water to satisfy our demands. Instead, it is a water management crisis wreaking havoc on billions of people and the environment.

## III. Facts Regarding Global Water Crisis

Every problem has some past evidence. The water crisis is not distinct from that.

**The 1700s to 1800s**: Increased urbanization in England resulted from industrialization, emphasizing the need for clean water supplies and sanitation.

**The 1800s**: The first documented water shortage in history.

**1854**: During the cholera outbreak in London, Dr. John Snow discovered the link between water and the transmission of the disease.

**The 1900s**: More than 11 billion people have perished due to drought since 1900, and more than one billion people have been afflicted.

**1993**: World Water Day was announced by the United Nations General Assembly on March 22.

**2000**: The United Nations' member states established Millennium Development Goals (MDGs) for progress in development, including a goal to reduce by half the number of people without access to clean drinking water by 2015.

**2003**: UN-Water was established as a coordination platform for sanitation and freshwater access challenges.

**2005**: Chronic water scarcity affects 35% of the global population, up from 9% in 1960.

**2005 to 2015**: In an International Decade for Action called "Water for Life," UN member nations prioritize water and sanitation development.

**2008**: The International Year of Sanitation, designated by the United Nations, places a high value on health and dignity.

**2010**: The MDG target for clean water availability was met five years early. Since 1990, almost two billion people have acquired access to clean drinking water. The United Nations General Assembly recognizes everyone's right to appropriate quantities of water for personal and domestic use that are physically accessible, equitably distributed, safe, and inexpensive.

**2013**: The United Nations has designated November 19 as World Toilet Day to draw attention to the global issue of billions of people who lack access to decent sanitation.

**2015**: In the previous 25 years, around 2.6 billion people have had access to safe drinking water, while approximately 1.4 billion have gained adequate sanitation access since 2000. The United Nations member states sign on to the Sustainable Development Goals (SDGs), which are the successors of the Millennium Development Goals (MDGs) and promise clean water and sanitation for all by 2030.

**2018**: Globally, 2.1 billion people still do not have access to safe drinking water in their houses, and over a billion people are forced to defecate outside.

## IV. Reasons Behind Water Crisis

The water crisis is a global concern. There are numerous reasons responsible for the water crisis. The main reasons for the water crisis are concisely explained in the following:

### A. Unequal Distribution of Water

The water crisis is closely related to the unequal distribution of water. In our society, the consumption of any natural resource follows an unequal distribution. In our society, neither equity nor equality is maintained for natural resource consumption in the same way as water consumption. Most of the time, unequal distribution of water use occurs, resulting in a water crisis. Due to the unequal distribution, only a limited number of individuals have access to specific resources. The unequal distribution also occurs globally. Therefore, it can also be called a global phenomenon. Developed countries consume high amounts of water, whereas underdeveloped or developing countries consume less water. Irrigation is highly hampered due to unequal distribution of water (Ariyonna, 2023). Several issues, including poverty, slow development, human migration, low GDP, corruption, a lack of social and economic fairness, and others, are linked to the unequal distribution of water that finally causes the water crisis.

### B. Water Pollution

One of the leading causes of the water crisis worldwide is pollution. Pollution can take many forms and almost always makes the water consumed by humans unsafe. Freshwater supply and reuse are increasingly threatened by water contamination (UNESCO, 2022). Water pollution indeed causes a shortage of safe water. Chemical or oil spills can permanently pollute the water. Poor sanitation and the lack of waste treatment plants are also responsible for the terrible pollution of water sources in rural areas (Rinkesh, 2022). Any industrial waste or feces dumped into rivers or oceans without proper treatment pollutes the water. If chemicals seep into groundwater or underground aquifers, pesticides and other fertilizers used by farmers can also cause water pollution. When contaminated groundwater enters the drinking water system, it poses a severe threat to public health and hampers the local economy. Water sources throughout the globe are being contaminated by oil spills. It pollutes the ocean and may infiltrate into freshwater supplies that people and other animals rely on for survival. It may leak into the water supply and contaminate the ocean through rivers, lakes, and other waterways (Chelsea, 2019). As a result, arsenic, asbestos, radon, agricultural chemicals, and hazardous waste in drinking water are increasing the risk of cancer (Vikramdeo et al., 2023).The World Bank stated in its new report, "Unknown Quality: The Invisible Water Crisis," that water pollution is already a problem. However, it will only deteriorate in the future due to global warming.

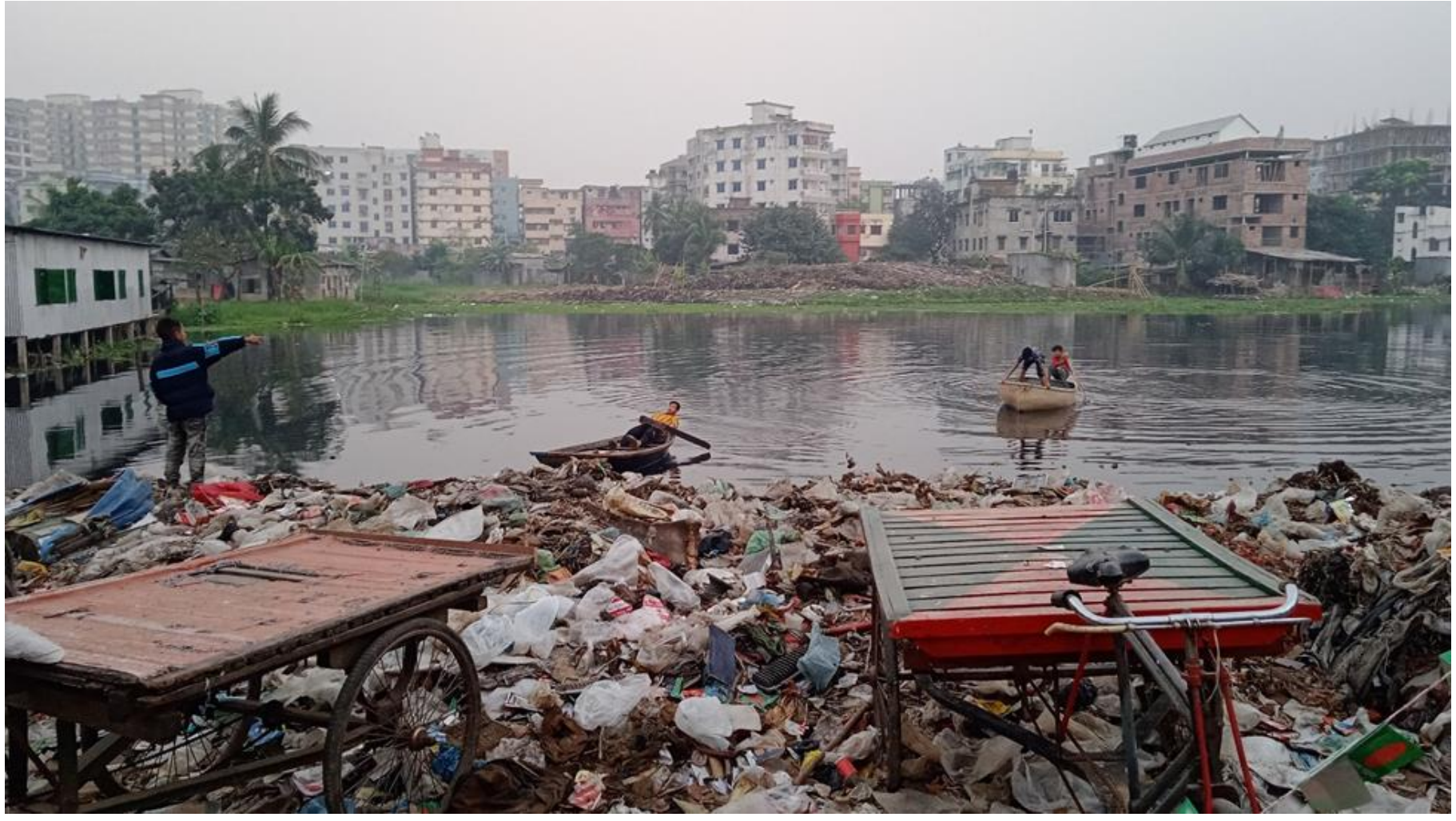

Fig. 1. Water pollution in Jhilpar lake, Dhaka, Bangladesh due to dumping of waste garbage.

As shown in Fig. 1, we have found a polluted lake in the Jhilpar area of Khilgaon Chowdhurypara, Dhaka, Bangladesh. Once upon a time, the lake was an amusement place for the local area people. Even many tourists visited here for enjoyment. But recently, some car workshops, tea stalls, grocery shops, etc., have been built near the Jhilpar lake. The waste garbages are dumped from the car workshops, tea stalls, grocery shops, etc., into the Jhilpar lake. It is now in critical condition. Formerly, people used the lake water for their daily needs, especially for drinking, cooking, bathing, etc., that's ehy they are affecting in skin and scalp diseases (Roy et al., 2022) but now they have to use groundwater, and to some extent, they have to buy water at high prices instead of using the lake water. Even some poor people use polluted lake water due to their insufficient income. There are no steps taken to control the waste garbage from Jhilpar lake. As a result, a water crisis extends in that region.

C. Overuse and Misuse of Water

Overuse and misuse of water are usually caused by human activities, mainly due to excessive groundwater use when the soil collapses, compacts, and drops. Excessive water pumping in coastal areas causes the saltwater to move inland and upwards, resulting in the water supply being contaminated by the saltwater. The overuse and misuse of water results in more water being wasted and squandered for insignificant reasons that increase the water crisis. During the drought, some people realized the value of water. But in recent years, the water crisis has become more than a temporary problem driven by the overuse and misuse of water. Now it is becoming systemic (Letmathe, 2014). The misuse and shortage of water nowadays are often to the detriment of the ecosystem, drying out lakes and rivers.

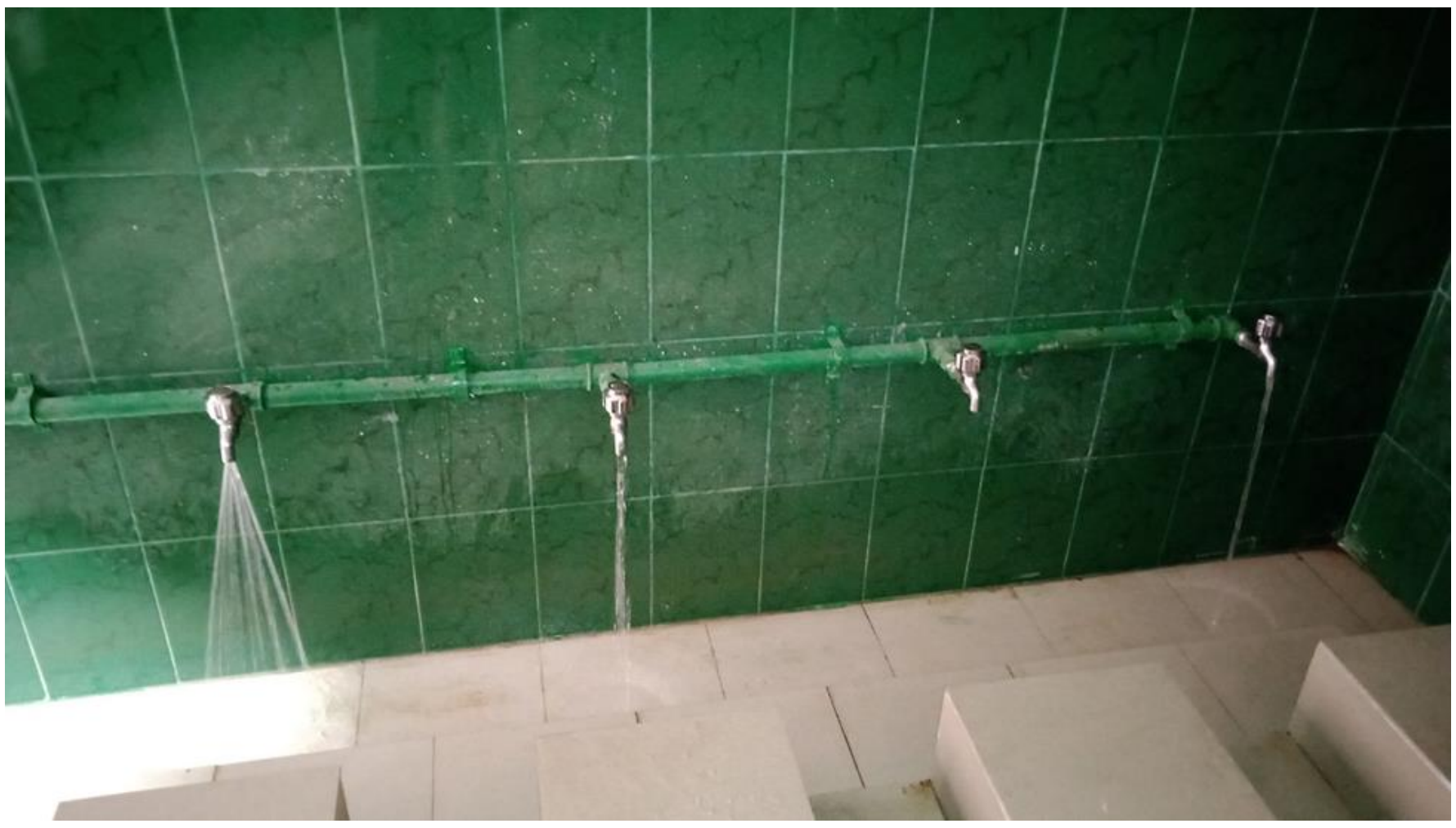

Fig. 2. Misuse of water in the ablution place of a mosque in Khilgaon, Dhaka, Bangladesh.

In Fig. 2, we have discovered the misuse of water in the ablution place of a mosque. Many people do wudu in the ablution place of the mosque before saying the prayer. After completing the wudu, some people came out without stopping the water tap. Even some people do wudu and refresh numerous times. A diverse range of people visit mosques, and some are ignorant of the importance of water. They are too illiterate to visualize the wastage of water and think about the water crisis. In many cases, sometimes there is no water in the water tap in the washrooms. Many people turn on the tap. When they find that there is no water in the tap, they either forget to turn off the water tap or leave without turning it off. Then when the water comes, and when there is no one around, the water falls unnecessarily and becomes wasted. A huge amount of water is wasted due to the improper activities of some people. In many cases, people do not turn off the water taps properly, resulting in unnecessarily wasting a small amount of water. This is also a kind of wastage of water, due to which a water crisis can be seen in the surrounding areas.

In the ablution place of the mosque, mainly groundwater is used for ceremonial washing before prayer and other purposes. It is found that there is a massive amount of misuse of water in the ablution places of mosques due to the illiteracy and irresponsibility of the people. The shortage of groundwater is often noticed due to the water abuses in mosques. There is a lack of administration in controlling the misuse of water in mosques. Such a large amount of water wastage frequently leads to a groundwater crisis, but we cannot recognize or solve the problem. For this reason, many people have to face the water crisis.

D. Mismanagement of Water Consumption

Mismanagement of the resource has resulted from rapid urbanization and unregulated water consumption. Water is considered an open-access resource, resulting in overuse and uncontrolled groundwater removal. Improper training and education result in the waste of safe, clean water on a daily basis and the misuse of it in regions where it is not required. Despite having a large population and a diversified terrain and climate, most developing countries lack a coherent water strategy. There are no strict guidelines for how various sectors and states should use surface and groundwater (Rinkesh, 2022). Activists and scholars have widely addressed the privatization of water.

E. Groundwater Depletion

Groundwater depletion occurs due to frequent water extraction from the ground. Most people pump groundwater continually from aquifers, and it does not have sufficient time to refill. Using a high volume of groundwater to meet our daily demands and irrigation depletes the groundwater. Continuous groundwater pumping is determined as the primary cause of groundwater depletion. Groundwater depletion has several negative consequences, including well dryness, less water in streams and lakes, water quality degradation, higher pumping costs, and land subsidence (USGS, 2018). Sometimes it also occurs naturally due to the variation of weather patterns, but it frequently occurs due to anthropogenic activities. We use groundwater for irrigation, domestic, industrial, and other purposes. If a high volume of water is extracted, it often creates a crisis for water.

F. Lack of infrastructure

It is merely the beginning of having adequate water to go around. But the water must be transported, treated, and disposed of. Waterborne and sewage-related pathogenic germs, viruses, parasites, and toxic chemicals threaten human health, and effective water infrastructure systems protect them. Unfortunately, global water infrastructure is deteriorating globally, such as treatment facilities, pipelines, and sewer systems (Schleifer, 2017).

G. Lack of institutions

Understanding the situation of water governance is crucial, since the water crisis is an outcome of the governance and analysis of water institutions (Ashish & Parthasarathy, 2021). Water institutions play a vital role in managing water resources. However, underdeveloped countries lack the institutional capacity to provide water treatment and management advice. Lack of institutions results in mismanagement and waste of water. Finally, it has become a water crisis.

H. Unutilized resources

In evaluating the economic cost of initiatives for developing water resources, the existence of underutilized resources has been identified as essential. Watersheds, catchments, and river basins have not been used effectively to protect water and soil. As a result, it influences river basin hydrology (Rinkesh, 2022).

I. Climate Change

Climate change disrupts weather patterns and leads to extreme weather events, unexpected water availability, an exacerbation of water shortages, and water pollution (UNICEF, 2023). The quality and quantity of water required for human survival may have serious consequences. Nowadays, climate change is felt mainly by water changes. Extreme weather events and variations in water cycle patterns exacerbate barriers to safe drinking water for the most vulnerable people. Rising temperatures in the freshwater source can lead to deadly pathogens, causing humans to consume harmful water. Rising sea levels cause salty freshwater to become salty, and millions of people threaten water resources (UNICEF, 2023). Climate change changes the way water evaporates and rains, and in both hemispheres, it pushes the precipitation further south. The pattern of rainfall has been significantly altered due to global warming.

J. Human Settlements

Population growth, climate change, and aging urban infrastructure are just a few of the stresses that cities throughout the globe are feeling right now. Many cities may confront difficulties efficiently managing scarcer and less dependable water supplies as future water demand rises. Ground realities and the difficulties of anticipated demands have made it clear that continuing with business as usual is not a viable option (UNESCO, 2022). Water issues are as diverse as the human settlements that depend on water accessibility.

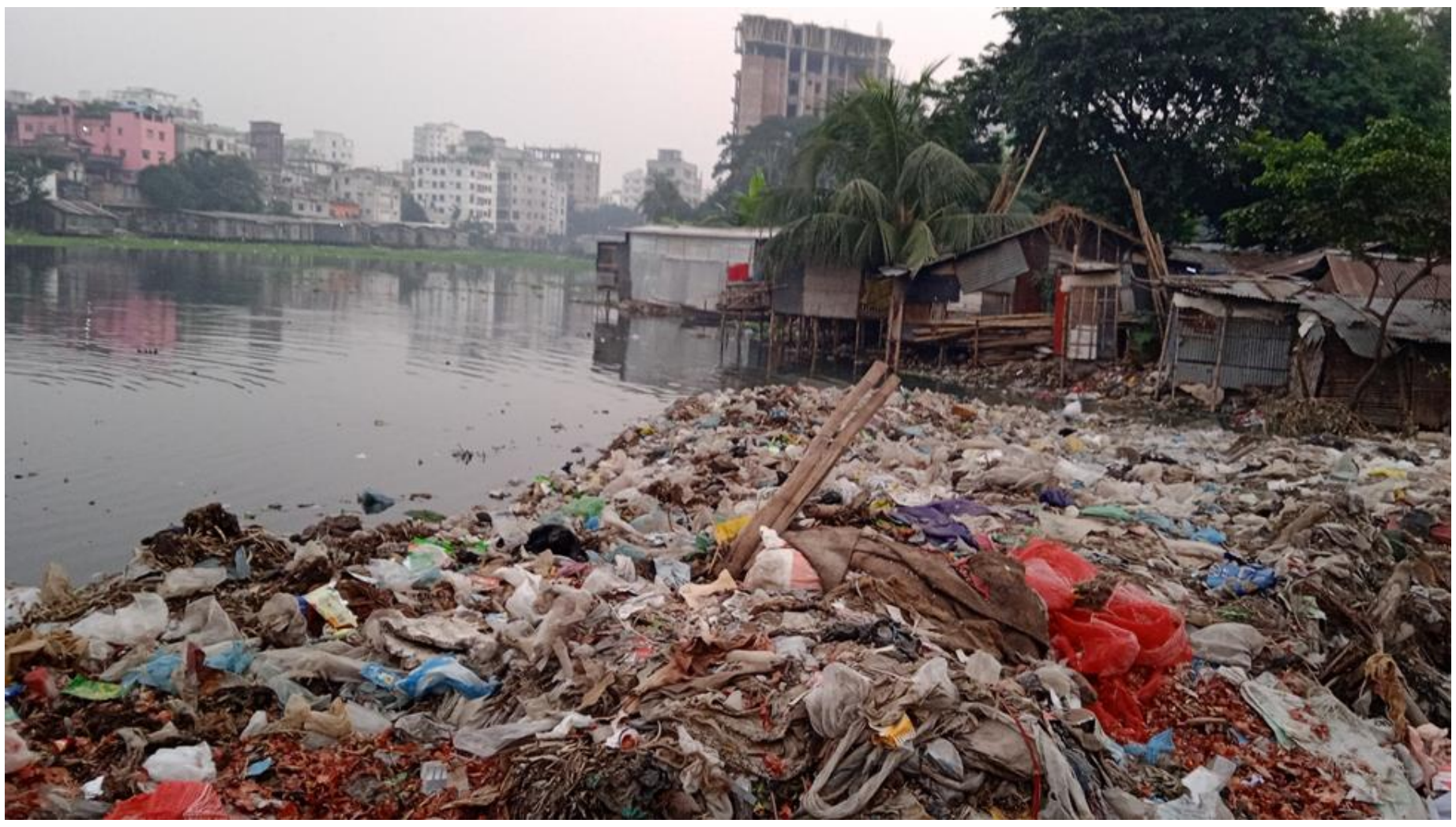

Fig. 3. Human settlements and waste dumping near Jhilpar lake, Dhaka, Bangladesh.

Due to human settlements, water demands and water pollution are both increasing. Large river ecosystems have been deliberately destroyed due to dams, hydroelectric projects, and water diversion for agriculture. Human settlements increase the water footprint that is finally creating the water crisis. In large areas such as Asia, the number of water-related issues is vast enough to respond to any crisis, whether flood or drought, pollution or displacement, not to reimburse tribute for water. Human settlements are being enhanced as population growth increases day by day. Due to human settlements, water demands and water pollution are both increasing. Large river ecosystems have been deliberately destroyed due to dams, hydroelectric projects, and water diversion for agriculture. Human settlements increase the water footprint that is finally creating the water crisis.

In Fig. 3, we have observed many houses built near the Jhilpar lake in Khilgaon chowdhurypara, Dhaka, Bangladesh. The authorities do not permit people to build houses adjacent to the lake. But due to a lack of monitoring from the administration, people build houses in the open spaces near the lake and pollute the lake water by dumping their household waste. The huge amount of household waste is mixed with the lake water and pollutes the lake water. There are also some shops, workshops, etc., that are also responsible for the Jhilpar lake water pollution. Therefore, the Jhilpar lake water cannot be used for human requirements. Once, this Jhilpar lake water was a good source of human water requirements. But due to the pollution, all of the people here are dependent on groundwater. When there is a water shortage, they buy water from WASA. Due to insufficient income, some of them have to use polluted lake water to fulfill their demands and suffer from various waterborne diseases. Thus, the water crisis occurs in this area due to lake water pollution.

### K. Corruption

Water corruption is the catalyst for the global water crisis. The corruption of essential resources and services threatens human life and development. The governance crisis in the water sector is mainly due to corruption. Corruption in water resource management hampers the long-term viability of water supplies, generates extremely uneven water distribution that can lead to political conflict, and fosters the destruction of vital ecosystems. Political people often use their political powers to consume much more water. Sometimes, people reserve vast amounts of water, and when there is a water shortage, they sell it at high prices. Reservations of water often create a crisis of water. When corruption occurs in the water sector, it diverts progress for everyone. It hinders investment, inhibits adequate water resource management and service supply, and degrades governmental institutions, leading to unsustainable development. Futhermore, it disproportionately affects the lower class, creating poverty by diminishing efficacy and efficiency (Stalgren, 2006).

### L. Unfair Pricing on Water

Water is grossly undervalued across the world. Transportation costs, treatment costs, and disposal costs are not included into the pricing. As a result, there has been a waste of water due to improper allocation and a failure to invest in water-saving infrastructure and technology (Schleifer, 2017). Moreover, the rich people might have the ability to buy potable water at high rates. However, the middle class and poor people are often unable to purchase potable water at inflated prices (Roy, Protity, Das, & Dhar, 2023). People often have to pay exorbitant prices in extreme poverty areas to get potable water. Those without money have to drink from holes in the ground or puddles by the roadside (Rinkesh, 2022). Therefore, they cannot ensure proper hygiene and suffer from various diseases.

## V. Potential Health Effects of the Current Water Crisis

Nwidu, Oveh, Okoriye, & Vaikosen (2008) conducted a three-year retrospective study on 100 participants from 2005 to 2007, including males, females, and children, in the Amassoma general hospital in order to evaluate the incidence of waterborne diseases in Nigeria. The study revealed that cholera, diarrhea, dysentery, and typhoid were the four most prevalent waterborne diseases among the local population. Over the course of the three-year study, the incidence of diarrhea (44.94%), dysentery (16.85%), and typhoid (34.83%) increased, while the incidence of cholera (3.37%) decreased over time shown in Fig. 4. They found that the contaminated water of the Amassoma river is used for the household chores and it is the main reason for the prevalence of water borne diseases (Nwidu, Oveh, Okoriye, & Vaikosen, 2008).

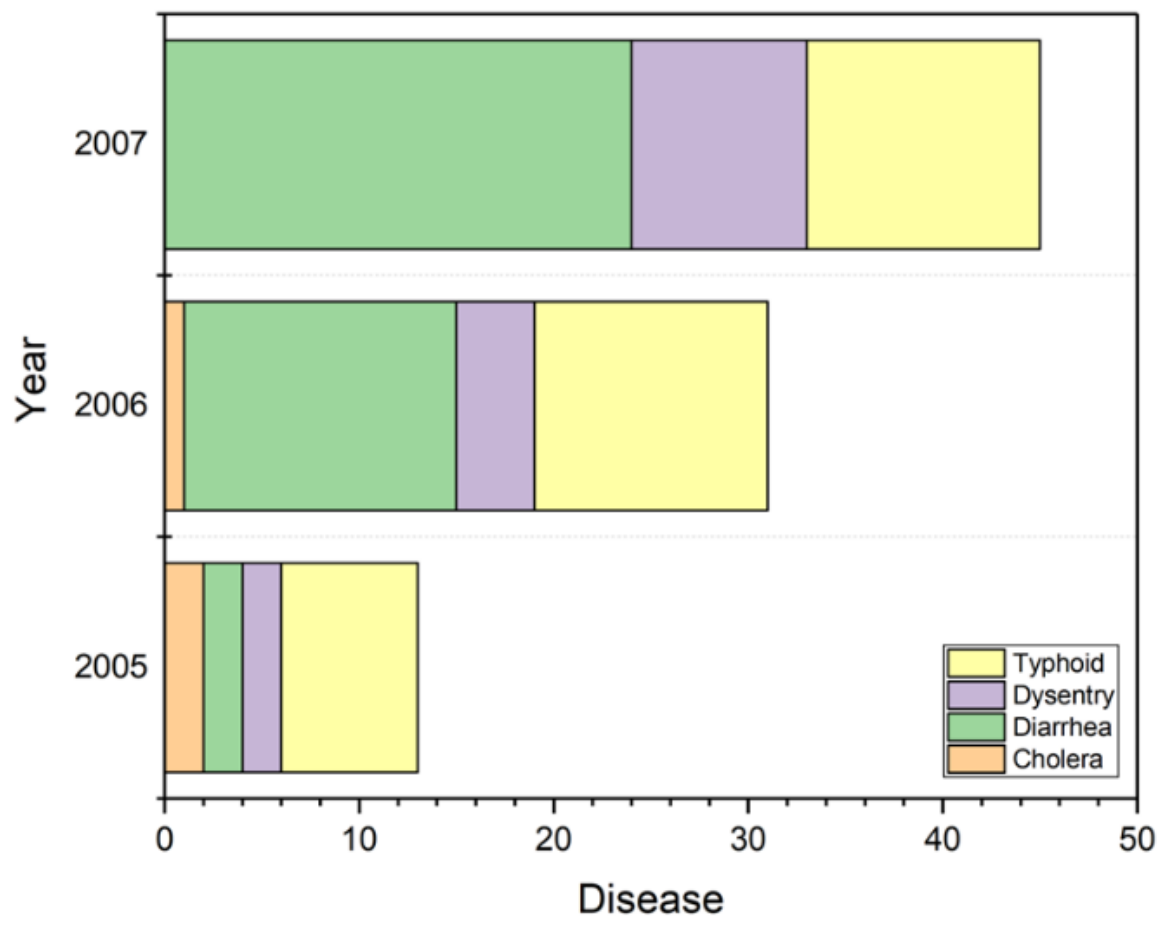

Fig. 4. Prevalence of cholera, diarrhea, dysentery and typhoid in Amassoma, Nigeria during the year 2005 to 2007.

The rapid and hazardous transmission of diseases can be attributed to the inherent solubility nature of water. In many low- and middle-income nations, waterborne parasite infections represent a prominent and consequential public health and economic challenge (Ngowi, 2020). Factors like the presence of harmful microorganisms such as bacteria, viruses, and parasites, as well as unsanitary conditions that can lead to the contamination of water sources. In addition, vectors such as insects or animals can contribute to the spread of waterborne illnesses. Moreover, the presence of toxic substances such as chemicals and heavy metals can also pose a significant risk to public health when present in water sources (Kulinkina, Shinee, Herrador, Nygard, & Schmoll, 2016). According to Global Infectious Disease and Epidemiology Online Network (GIDEON) database search, from 2000 to 2013 total 1039 outbreaks were recorded among which 185 (18%) were specifically linked to water. The 185 water-based outbreaks covered 18 diseases shown in Fig. 5.

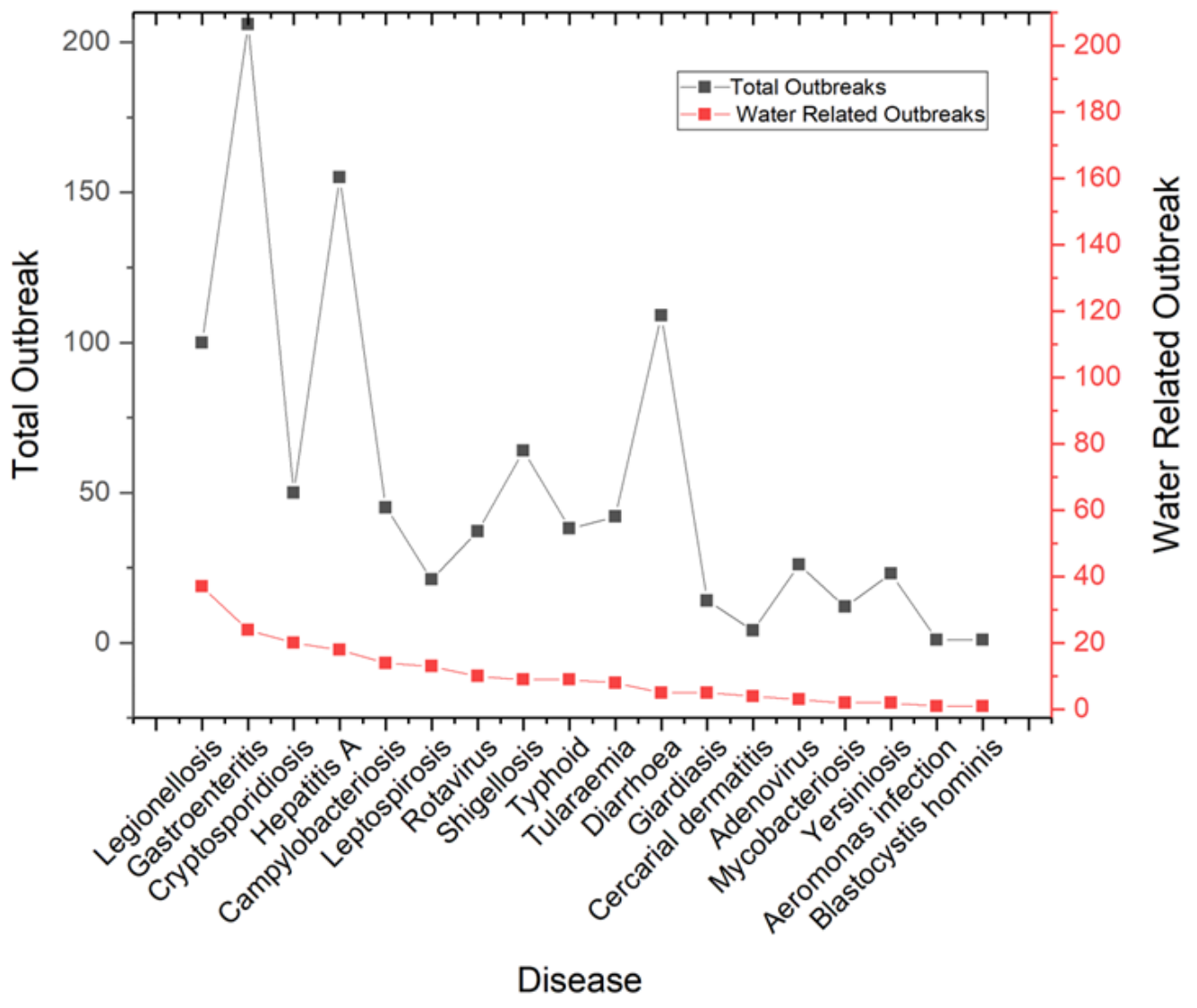

Fig. 5. Outbreaks reported in GIDEON between 2000 to 2013.

According to B. J. Pauli et. al. (Meehan, et al., 2020), in 2014 and 2015, Flint and the surrounding Genesee County saw a historical outbreak of legionnaires' disease, a severe form of pneumonia. According to their research, the death toll was 13 or maybe significantly higher due to legionnaires' disease. Despite the Michigan health department's attribution of blame to local hospitals' ignorance, multiple studies strongly support the notion that the increase in household infections during the summer of 2014 was mainly due to the water switch (Meehan, et al., 2020).

Motoshita, et. al. (2011) performed a regression study to look at the connection between water scarcity and infectious illnesses. The results indicated that Asia and Africa are more likely to experience diarrhea and other intestinal nematode diseases. The investigation came to the conclusion that the lack of access to clean water for residential use is the primary cause of the onset of these diseases (Motoshita, Itsubo, & Inaba, 2011).

M. A. Borchardt, et. al. (2012) worked to quantify the burden of waterborne acute gastrointestinal illness (AGI) due to untreated groundwater in Wisconsin. 580 adult and 1,079 children participated in the survey and 1,204 tap-water samples were collected from 14 communities. Between 6% and 22% of AGI was thought to be caused by viruses (Adenovirus, Enterovirus, Hepatitis A virus, Rotavirus, norovirus GI, norovirus GII) found in tap water. They also observed that the tap-water virus most strongly linked to the occurrence of AGI in the areas was norovirus GI (Borchardt, Spencer, Kieke, & Lambertini, 2012).

Water is the home of vector borne diseases. One concerning vector borne outbreak is malaria, a non-communicable disease (Roy, Protity, Das, & Dhar, 2023). The tropical form of malaria, Plasmodium falciparum, which is prevalent in the tropics and sub-Saharan Africa, is responsible for about 90% of malaria cases worldwide and causes the most severe clinical form of the disease (Caminade, McIntyre, & Jones, 2018). A minimum air temperature of roughly 15-20°C is required for the parasite to replicate within the mosquito vector, and as daily temperatures fluctuate closer to this threshold, transmission tends to rise (Martens, Jetten, & Focks, 1997). The creation of ample vector breeding grounds is facilitated by precipitation, while the correlation between temperature, rainfall, and malaria is also largely dependent on the specific vector involved (Hope, A., Hemingway, & McKenzie, 2009). These simulations show that the next century's climate will become more favorable to malaria transmission in tropical elevated regions, particularly in East Africa (Gething, et al., 2010). The complexity of the malaria parasite has made it difficult to develop effective vaccines, though the most advanced candidate, RTS,S/AS01, is set to begin clinical trials in three African countries in 2018. Transmission-blocking immune responses to the malaria parasite are known to exist (Gething P. W., et al., 2010).

Despite the geography of Nepal being primarily mountainous and being at least 2000 m above sea level, there were 2,092 confirmed cases of malaria-related deaths in 2012 (Dhimal, Ahrens, & Kuch, 2015). They also found the evidence of four other water-borne vector-borne diseases, lymphatic filariasis, Japanese encephalitis, visceral leishmaniasis, and dengue outbreaks occurred from 1970 until 2014 in Nepal (Dhimal, Ahrens, & Kuch, 2015). Over the past five decades, there has been a substantial 30-fold surge in the worldwide incidence rate of dengue, making it the most rapidly disseminating mosquito-borne illness known to date (Hemingway, 2017).

## VI. Conclusion

The water problem significantly affects public health, especially in developing nations. A lack of access to sanitary facilities and safe drinking water facilitates the spread of waterborne diseases and other health issues. Additionally, the water issue has worsened due to the overuse and depletion of natural water sources brought on by the growing population and industrialization. Collaboration between governmental bodies, non-governmental groups, and individuals is necessary to address the water situation. The effects of the water crisis on public health can be lessened by implementing sustainable water management practices, investing in infrastructure and technology, and fostering awareness and education. Unless immediate and concerted action is taken, the water crisis will persist, posing a severe risk to public health worldwide.

## VII. Acknowledgment

The authors would like to express their heartfelt gratitude to Almighty Allah for the accomplishment of this research.

## VIII. Funding

Self-funded.

## IX. Conflict of Interest

The authors willingly declare that they have no potential of interest for the publication of this research findings.